\documentclass{article}

\usepackage[a4paper, margin={1.5in}]{geometry}
\usepackage{hyperref}
\usepackage{amsmath}
\usepackage{amssymb}
\usepackage{booktabs}
\usepackage{listings}
\usepackage{xcolor}
\usepackage{xspace}
\usepackage{tikz}
\usetikzlibrary{arrows.meta,positioning}

\lstdefinestyle{pacmplstyle}{
    basicstyle=\ttfamily\footnotesize,
    keywordstyle=\bfseries,
    commentstyle=\itshape\color{gray},
    stringstyle=\color{purple},
    numbers=left,
    numberstyle=\tiny\color{gray},
    stepnumber=1,
    numbersep=5pt,
    showspaces=false,
    showstringspaces=false,
    breaklines=true,
    captionpos=b              
}

\lstdefinelanguage{concepts}{
  morekeywords={concept, purpose, state, actions, requires, effects, application, types, concepts, invariants, synchronizations, reaction, when, where, then, action, error, ok, nok}, 
  sensitive=true,                                             
  morecomment=[l]{//},                                        
  morecomment=[s]{/*}{*/},                                    
  morestring=[b]",                                            
  morestring=[d]'                                             
}

\newcommand{\inlineconcept}[1]{\lstinline[language=concepts, style=pacmplstyle]{#1}}

\newcommand{\tool}{\textsc{foundry}\xspace}

\author{Alcino Cunha\\INESC TEC \& Universidade do Minho, Braga, Portugal}
\title{Verified LLM-Driven Synthesis for Concept Design}
\date{July 2026}

\begin{document}

\bibliographystyle{alpha}

\maketitle

\begin{abstract}
\emph{Concept Design}~\cite{jackson2021essence} structures software systems around \emph{concepts}: user-facing, self-contained units of functionality with a focused purpose. Concepts are composed into applications using synchronization rules called \emph{reactions}, which specify how actions in one concept trigger actions in others. This paper first gives a formal semantics for concepts and reactions, enabling automatic verification of safety invariants in applications developed with this methodology. It then presents a CEGIS-style, LLM-driven synthesis procedure for generating reaction designs that satisfy such invariants. Because many different designs can satisfy the same invariant, we study two ways of steering synthesis toward the user's intended design: natural-language prompts and positive/negative scenarios. We also propose an LLM-driven scenario elicitation technique to support early design exploration. In an evaluation on three applications and twelve design variants using one LLM configuration, invariant-only synthesis reached verified designs quickly but often produced inconsistent designs across runs, some of which were implausible, showing that invariants alone underconstrain the design task. Scenario-guided synthesis recovered intended designs more consistently than natural-language prompting, although minimal scenarios can lead to overfitting. LLM-driven scenario elicitation, where the user classifies proposed scenarios rather than authoring them from scratch, recovered the intended designs in most variants when enough scenarios were elicited, but missed behaviors and non-determinism prevented reliable coverage in all cases.
\end{abstract}

\section{Introduction}

Modern software is often assembled by composing reusable units of behavior: authentication, labeling, sharing, notification, access control, and many others. The difficult part is rarely defining the behavior of each unit in isolation, but specifying the coordination logic that makes the units work together safely. Concept Design~\cite{jackson2021essence} provides a structuring discipline for this kind of composition by organizing applications around reusable components called \emph{concepts}: user-facing, self-contained units of functionality that act as micro-applications with focused purposes. Concepts are assembled into complete applications using synchronization rules called \emph{reactions}, which specify when an action in one concept should trigger other actions in the same or a different concept.

Because reactions define the coordination logic between concepts, safety properties often depend not on any single concept in isolation, but on the way reactions dictate how concepts interact. For example, an application may need to ensure that access-control roles are synchronized with the enrollment status, or that shared evidence is decrypted only while a valid sharing link exists. Such properties are difficult to validate informally, especially if the application includes many reactions. Yet Concept Design does not currently provide a formal semantics for reactions, which makes it hard to state precisely what a design means, to verify that it preserves a desired invariant, or to synthesize safe coordination logic. Our first contribution is therefore to give an automata-based semantics for concepts and reactions. The semantics is deliberately simple: it can be encoded in high-level formal specification languages such as Alloy 6~\cite{jackson2012software,macedo2016lightweight} or TLA+~\cite{lamport1994temporal}, and it supports automatic checking of both safety invariants and scenarios.

Once designs can be checked automatically, they can also be synthesized automatically: an LLM can propose reaction candidates, a verifier can check if those candidates satisfy the desired invariant, and give back to the LLM counterexamples to help refine the proposed reactions if they do not. This \emph{Counterexample-Guided Inductive Synthesis} (CEGIS)~\cite{solar2008program} loop is rather straightforward, but it exposes a deeper problem. Verification can tell us whether a design satisfies the safety invariant; it does not tell us whether the design is the one the user intended. In practice, many different reaction designs can satisfy the same invariant, including designs that repair violations in undesirable ways or forbid behaviors that the user expected to allow. Intent often lives outside the safety invariant the designer is willing or able to formalize.

To narrow this design space, we propose using positive and negative scenarios as checkable steering artifacts. A positive scenario describes a behavior that the synthesized design should permit; a negative scenario describes one that it should reject. Unlike natural-language prompts, scenarios have a semantics: a candidate design can be checked against them automatically. This lets scenarios play two roles at once. They guide the LLM toward the intended design, and they provide additional formal checks that reject designs whose behavior disagrees with the user's intent. We realize this idea in \tool, a prototype that uses LLMs for reaction synthesis and translates concept designs to Alloy 6 for verification. Scenarios still require effort to write, and early in design the user may not even have a clear intended design. To support this phase, \tool also includes an LLM-driven scenario elicitation procedure. The model proposes short invariant-violating prefixes and possible repairs; the user classifies them as desirable or undesirable; and the resulting classified scenarios can then be fed to the synthesis loop.

Although our setting is Concept Design, the problem is broader: synthesizing coordination logic among reusable behavioral components under safety constraints. We evaluate \tool on three applications and twelve design variants using one frontier LLM configuration, namely GPT-5.5. The results show that invariant-only synthesis often reaches verified designs quickly, but rarely converges on a stable design, and can even produce implausible designs. Scenario guidance recovers intended designs more consistently than natural-language prompting in our benchmark, while LLM-driven scenario elicitation shifts part of the specification task from authoring scenarios to classifying proposed ones, although coverage gaps and overfitting remain important failure modes.

In summary, this paper makes the following contributions:
\begin{itemize}
    \item An automata-based semantics for concepts and reactions, enabling
    bounded verification of invariants and scenarios.
    \item A verified LLM-driven synthesis loop for reaction designs, together with a scenario-guided mechanism for steering synthesis toward intended designs.
    \item An LLM-driven scenario elicitation procedure for exploring design
    alternatives.
    \item An evaluation on three applications and twelve design variants, using one LLM configuration, comparing scenario-guided synthesis with natural-language prompting and characterizing recurring failure modes.
\end{itemize}

The remainder of the paper is organized as follows. Section~\ref{sec:overview} introduces Concept Design, reactions, verification, and synthesis with \tool through a running example. Section~\ref{sec:semantics} presents the automata-based semantics of concepts and reactions, and shows how invariants and scenarios can be formally checked. Section~\ref{sec:synthesis} describes the LLM-driven CEGIS-style synthesis loop and the scenario elicitation procedure implemented in \tool. Section~\ref{sec:evaluation} evaluates the approach on three applications and twelve design variants. Section~\ref{sec:relatedwork} discusses related work, and Section~\ref{sec:conclusion} concludes.

\section{Overview}
\label{sec:overview}

Listings~\ref{lst:label} and~\ref{lst:dlq} present two concepts, described in a lightweight DSL similar to the one introduced in \cite{jackson2021essence}. The first is a common concept in many applications: labeling items with tags. Its state consists of a single association between items and their respective sets of tags, and it provides three actions described in the classic pre- and post-condition form. The second is a \emph{Dead-Letter Queue} (DLQ), a concept used in production systems to retain messages that could not be processed so that they can later be retried or debugged. Its state consists of two sets of messages: one containing pending messages that have not yet been processed, and another containing the ``dead'' messages whose processing failed. Its behavior comprises five actions, one of which purges all dead messages.

\begin{lstlisting}[float, language=concepts, style=pacmplstyle, caption={A labeling concept}, label={lst:label}]
concept: Label[Item,Tag]
purpose: To allow labeling items with tags.
state:
    labels: Item -> set Tag
actions:
    action affix(i:Item, t:Tag)
        requires: t is not a label of i
        effects:  adds t to the labels of i
    action detach(i:Item, t:Tag)
        requires: t is a label of i
        effects:  removes t from the labels of i
    action clear(i:Item)
        requires: i has at least one label
        effects:  removes all labels from i
\end{lstlisting}

\begin{lstlisting}[float, language=concepts, style=pacmplstyle, caption={A dead-letter queue concept}, label={lst:dlq}]
concept: DeadLetterQueue[Message]
purpose: To process messages while isolating failed messages for later retry.
state:
    pending: set Message
    dead: set Message
actions:
    action submit(m:Message)
        requires: m is not in pending or dead
        effects:  adds m to pending
    action succeed(m:Message)
        requires: m is in pending
        effects:  removes m from pending
    action fail(m:Message)
        requires: m is in pending
        effects:  moves m from pending to dead
    action retry(m:Message)
        requires: m is in dead
        effects:  moves m from dead to pending
    action purge()
        requires: dead is not empty
        effects:  removes all messages in dead
\end{lstlisting}

Listing~\ref{lst:sensitive} presents an application that composes these two concepts. Concepts are composed using \emph{synchronizations}: rules that describe how the system reacts to occurrences of actions in one concept by invoking actions in other concepts. The DSL for describing these synchronization rules has evolved since the Concept Design theory was first presented in~\cite{jackson2021essence}. Here, we use a style similar to the latest version of that DSL, as presented in~\cite{meng2026making}. Each synchronization rule is a \emph{reaction} that fires \emph{when} a specific action occurs in a concept and \emph{where} a specific condition on the state is met. The rule \emph{then} triggers a (re)action in another (or the same) concept. Note that the triggered action can in turn fire other synchronization rules, thus enabling complex causal chains of reactions.

In critical applications, safety invariants often need to be enforced. In this example, messages can be labeled with a special tag indicating that they contain sensitive data. Since the dead messages in a DLQ are often copied into less protected operational storage, we require that no dead message be labeled as sensitive. Listing~\ref{lst:sensitive} defines two reactions that enforce this safety requirement: the first states that if the sensitive tag is affixed to a ``dead'' message, that message should be retried, so that it no longer remains in the DLQ; the second states that marking a sensitive message as failed should not be permitted, and is therefore handled by triggering the special \inlineconcept{error} action.

\begin{lstlisting}[float, language=concepts, style=pacmplstyle, caption={A dead letter queue with sensitive data}, label={lst:sensitive}]
application: NoDeadSensitiveMessages
types:
    a set Message
    a set Tag
    one special Tag named Sensitive
concepts:
    one DeadLetterQueue[Message] named Q
    one Label[Message,Tag] named L
invariants:
    there are no dead messages in Q labeled with Sensitive in L
synchronizations:
    reaction retry_sensitive_dead_message
        when  L.affix(m,Sensitive)
        where m is in dead in Q
        then  Q.retry(m)
    reaction prevent_sensitive_failure
        when  Q.fail(m)
        where Sensitive is a label of m in L
        then  error
\end{lstlisting}

One of the capabilities of \tool is to formally verify that an application's design (its set of reactions) ensures the desired invariant. It does so by translating both the concepts and the reactions into the Alloy~\cite{jackson2012software,macedo2016lightweight} formal specification language, using the semantics presented in the next section, and then invoking the Alloy model checker~\cite{brunel2018electrum} to perform the verification. \tool can also check that the application does not over-react: that is, that it does not react (in particular, with an error) in states where the invariant already holds. For example, if we remove the first reaction and verify the resulting design, \tool returns the following counterexample.
\begin{lstlisting}[style=pacmplstyle, numbers=none]
Sequence of actions:
  1. Q.submit(Message1)
     ongoing reactions: (none)
  2. Q.fail(Message1)
     ongoing reactions: (none)
  3. L.affix(Message1, Sensitive)
     ongoing reactions: (none)
After this sequence of actions the invariant is broken but the app is not reacting
\end{lstlisting}
Conversely, if we remove the \emph{where} condition from the first reaction, we obtain the following counterexample, showing that the application now reacts even in a situation where the invariant is not broken.
\begin{lstlisting}[style=pacmplstyle, numbers=none]
Sequence of actions:
  1. L.affix(Message0, Sensitive)
     ongoing reactions: retry_sensitive_dead_message(Message0)
After this sequence of actions the invariant is valid but the app is reacting
\end{lstlisting}

Leveraging this verification component, the main functionality of \tool is to synthesize designs using a CEGIS-style agentic loop, in which an LLM proposes reactions that are then verified against the invariant. If the design fails verification, the counterexample is fed back to the LLM to guide refinement; this process repeats until a correct application design is obtained or a user-defined maximum number of iterations is reached. Given the desired invariant alone, for simple applications, such as the one above, GPT-5.5 can often synthesize a correct design on the first attempt. For example, when we passed a partial description of this application (without any reactions) to \tool, GPT-5.5 returned the following correct set of reactions.
\begin{lstlisting}[language=concepts, style=pacmplstyle, numbers=none]
synchronizations:
    reaction remove_sensitive_label_on_failure
        when  Q.fail(m)
        where Sensitive is a label of m in L
        then  L.detach(m,Sensitive)
    reaction retry_dead_message_on_sensitive_label
        when  L.affix(m,Sensitive)
        where m is in dead in Q
        then  L.detach(m,Sensitive)
\end{lstlisting}

Although this design satisfies the desired invariant, it is implausible and most likely not one a user would want, since it automatically declassifies sensitive messages that fail to be processed or dead messages that have been classified as sensitive. For a given invariant, many different designs can ensure it. To guide the synthesis process toward the intended design, \tool offers a scenario-guided synthesis loop in which the user provides a few guiding positive and negative scenarios. For example, if the user provides the following two scenarios, GPT-5.5 synthesizes the design in Listing~\ref{lst:sensitive} on the first attempt.
\begin{lstlisting}[style=pacmplstyle, numbers=none, language=concepts]
nok: Q.submit(m); L.affix(m,Sensitive); Q.fail(m)
ok:  Q.submit(m); Q.fail(m); L.affix(m,Sensitive); Q.retry(m)
\end{lstlisting}

In the initial phases of the design, the user often does not have a clear idea about the intended design, so they cannot yet provide the necessary scenarios. To help kickstart the design process, \tool also offers an option whereby an LLM proposes some scenarios and asks the user to classify them. Using this option, if we ask GPT-5.5 to generate five scenarios for our application, it proposes the following.
\begin{lstlisting}[style=pacmplstyle, numbers=none, language=concepts]
L.affix(m,Sensitive); Q.submit(m); Q.fail(m);
Q.submit(m); Q.fail(m); L.affix(m,Sensitive);
Q.submit(m); Q.fail(m); L.affix(m,Sensitive); Q.retry(m);
L.affix(m,t); Q.submit(m); Q.fail(m); L.affix(m,Sensitive);
L.affix(m,t); Q.submit(m); Q.fail(m); L.affix(m,Sensitive); Q.retry(m);
\end{lstlisting}
If the user classifies the first as undesirable and all the others as desirable, and we add these classifications to the (partial) application description, \tool with GPT-5.5 can also synthesize the design in Listing~\ref{lst:sensitive} in very few iterations.

\section{Formal Verification of Concept Designs}
\label{sec:semantics}

This section defines the transition-system semantics that \tool uses to interpret concept designs and reduce invariant and scenario checking to model checking of temporal logic formulas.

\subsection{Concept Semantics and Composition}

When Concept Design was first proposed~\cite{jackson2021essence}, concepts were modeled as deterministic state machines. Here we refine this semantics slightly and model each concept by an \emph{initialized partial semiautomaton} (or simply, an \emph{semiautomaton}), a deterministic state machine with inputs but no outputs, where the inputs correspond to the set of actions the automaton can execute (a \emph{Mealy machine} without outputs). Formally, a concept is a tuple $C = (S,\Sigma,\delta,i)$, where $S$ is the set of states, $\Sigma$ the set of inputs (actions), $\delta : S \times \Sigma \rightharpoonup S$ is the transition function, and $i \in S$ is the initial state.

For example, consider the \inlineconcept{Label} concept as used in the \inlineconcept{NoDeadSensitiveMessages} application. Given a set of messages $\mathsf{Msg}$ and a set of tags $\mathsf{Tag}$ (such that $\mathsf{sensitive} \in \mathsf{Tag}$), it is defined by the semiautomaton $C_\mathsf{Label} = (S_\mathsf{Label},\Sigma_\mathsf{Label},\delta_\mathsf{Label},i_\mathsf{Label})$, where
\begin{itemize}
    \item $S_\mathsf{Label} = \mathcal{P}(\mathsf{Msg} \times \mathsf{Tag})$
    \item $i_\mathsf{Label} = \emptyset$
    \item $\Sigma_\mathsf{Label} = \{\mathsf{affix},\mathsf{detach}\} \times \mathsf{Msg} \times \mathsf{Tag} \cup \{\mathsf{clear}\} \times \mathsf{Msg}$
    \item $\delta_\mathsf{Label}$ is defined as follows.
    \begin{displaymath}
        \begin{array}{rcll}
            \delta(s,(\mathsf{affix},m,t)) & = & s \cup \{(m,t)\} & \text{if}\ (m,t) \not\in s \\
            \delta(s,(\mathsf{detach},m,t)) & = & s \setminus \{(m,t)\} & \text{if}\ (m,t) \in s\\
            \delta(s,(\mathsf{clear},m)) & = & s \setminus (\{m\} \times \mathsf{Tag}) & \text{if}\ \exists t . (m,t) \in s
        \end{array}
    \end{displaymath}
\end{itemize}

The concepts that operate in an application are composed using the standard parallel asynchronous composition. We assume that the set of inputs of every pair of concepts is disjoint, so the composition can be defined as

\begin{displaymath}
    C_1 \parallel C_2 = (S_1 \times S_2, \Sigma_1 \cup \Sigma_2,\delta,(i_1,i_2))
\end{displaymath} 
where
\begin{displaymath}
    \delta((s_1,s_2),a) = \left\{
    \begin{array}{rl}
        (\delta_1(s_1,a),s_2) & \text{if}\ a \in \Sigma_1 \text{ and } \delta_1(s_1,a) \text{ is defined}\\
        (s_1,\delta_2(s_2,a)) & \text{if}\ a \in \Sigma_2 \text{ and } \delta_2(s_2,a) \text{ is defined}
    \end{array}
    \right.
\end{displaymath}

\subsection{Reaction Semantics}

The reactions of an application can be modeled by a partial monitor semiautomaton that observes the composed concept automaton $C = (S, \Sigma, \delta, i)$ and tracks pending obligations.  Formally, a reaction 
\begin{displaymath}
    \mathsf{reaction}\ R\ \mathsf{when}\ W\ \mathsf{where}\ E\ \mathsf{then}\ T
\end{displaymath}
is well formed if $W$ and $E$ are first-order formulas whose conjunction defines a predicate over $S \times \Sigma$, and $T$ is a first-order formula that defines a predicate over $\Sigma$. The variables that occur in $T$ must also occur in the trigger condition, that is
\[
    \mathsf{fv}(T) \subseteq \mathsf{fv}(W \wedge E).
\]
Variables that occur in $W \wedge E$ but not in $T$ are local to the trigger condition and are existentially quantified when determining whether the reaction fires. The $\mathsf{error}$ predicate is an alias for $\bot$, the false predicate that never holds, meaning that an $\mathsf{error}$ obligation cannot be fulfilled and remains forever pending. In the formal model, an \inlineconcept{error} reaction marks the triggering situation as unrecoverable: once the obligation is created, no later action can discharge it, so the behavior can never return to a settled state. In an implementation, such a reaction should be realized by rejecting, guarding, or aborting the triggering action before its effects are committed.

Let $\mathcal{R}$ be the set of reactions in the application. For each reaction $R \in \mathcal{R}$, let $\Gamma_R$ be the set of all valuations of the variables in $\mathsf{fv}(T_R)$. Rather than giving each reaction a separate automaton, we define a single reaction monitor whose state is the set of all pending reaction instances. Since the same valuation may occur in different reactions, pending instances are tagged with the reaction that created them:
\[
    S_\mathcal{R} =
    \mathcal{P}\left(\bigcup_{R \in \mathcal{R}} \{R\} \times \Gamma_R\right).
\]
A monitor state $r \in S_\mathcal{R}$ contains a pair $(R,\Gamma)$ exactly when reaction $R$ has fired with valuation $\Gamma$ and the corresponding obligation has not yet been settled.

Given a concept state $s \in S$ and an executed action $\sigma \in \Sigma$, reaction $R$ fires for valuation $\Gamma \in \Gamma_R$ if there exists an extension $\Gamma'$ of $\Gamma$ to the variables in $\mathsf{fv}(W_R \wedge E_R)$ such that $(W_R \wedge E_R)[\Gamma'](s,\sigma)$ holds. The state $s$ is the source state of the concept transition, so the \emph{where} condition is evaluated before the effects of $\sigma$. A pending instance $(R,\Gamma)$ is settled when $T_R[\Gamma](\sigma)$ holds for the executed action $\sigma$.

Pending obligations have priority over other actions. We capture this by the following predicates:
\[
\mathsf{pending}(r) \iff r \neq \emptyset
\]
\[
\mathsf{resolves}(r,\sigma) \iff
\exists (R,\Gamma) \in r.\ T_R[\Gamma](\sigma).
\]

The reaction monitor is the partial semiautomaton
\[
    \mathcal{M}_\mathcal{R} =
    (S_\mathcal{R}, S \times \Sigma, \delta_\mathcal{R}, \emptyset)
\]
where
\[
\begin{array}{rcl}
\delta_\mathcal{R}(r,(s,\sigma)) &=&
    \left(r \setminus
    \{(R,\Gamma) \in r \mid T_R[\Gamma](\sigma)\}\right)
    \cup \\[0.25em]
&&  \{(R,\Gamma) \mid
        R \in \mathcal{R}
        \land
        \exists \Gamma'.\
        \Gamma'|_{\mathsf{fv}(T_R)} = \Gamma
	        \land
	        (W_R \wedge E_R)[\Gamma'](s,\sigma)\}.
\end{array}
\]
provided that $\neg\mathsf{pending}(r) \lor \mathsf{resolves}(r,\sigma)$, and undefined otherwise. Essentially, we first remove from the monitor state all pending obligations that were fulfilled and then add the new ones.

For example, consider the first reaction in Listing~\ref{lst:sensitive}.
When $\sigma = \mathsf{affix}(m,\mathsf{sensitive})$ and $m \in \mathsf{dead}(s)$, a pending obligation for that reaction is added to the monitor state for the valuation of $m$ that made the triggering condition true. When $\sigma = \mathsf{retry}(m)$ holds for the same valuation, that obligation is removed from the monitor state. In reaction \inlineconcept{prevent_sensitive_failure}, pending instances are never removed because the $\mathsf{error}$ predicate never holds.

\begin{figure}[t]
\centering
\resizebox{\linewidth}{!}{%
\begin{tikzpicture}[
    >={Latex[length=2mm]},
    statenode/.style={draw=black!45, rounded corners=1pt, align=left, inner sep=3pt, font=\scriptsize, minimum width=2.1cm},
    badnode/.style={statenode, draw=red!65!black, fill=red!8},
    trans/.style={->, thick, black!65},
    action/.style={font=\scriptsize, fill=white, inner sep=1pt, text=black}
]
\node[statenode] (s0) {$P=\emptyset,\ D=\emptyset$\\$L=\emptyset$\\$O=\emptyset$};
\node[statenode, right=2.35cm of s0] (p1) {$P=\{m_1\},\ D=\emptyset$\\$L=\emptyset$\\$O=\emptyset$};
\node[statenode, right=2.35cm of p1] (d1) {$P=\emptyset,\ D=\{m_1\}$\\$L=\emptyset$\\$O=\emptyset$};
\node[badnode, minimum width=3.2cm, right=2.35cm of d1] (d1l) {$P=\emptyset,\ D=\{m_1\}$\\$L=\{(m_1,\mathsf{sensitive})\}$\\$O=\{\mathsf{retry\_sensitive\_dead\_message}(m_1)\}$};
\node[statenode, right=2.35cm of d1l] (repaired) {$P=\{m_1\},\ D=\emptyset$\\$L=\{(m_1,\mathsf{sensitive})\}$\\$O=\emptyset$};

\node[statenode, below=1.55cm of p1] (l1) {$P=\{m_1\},\ D=\emptyset$\\$L=\{(m_1,\mathsf{sensitive})\}$\\$O=\emptyset$};
\node[statenode, right=2.35cm of l1] (l2) {$P=\{m_1,m_2\},\ D=\emptyset$\\$L=\{(m_1,\mathsf{sensitive})\}$\\$O=\emptyset$};
\node[statenode, right=2.35cm of l2] (l3) {$P=\{m_1\},\ D=\{m_2\}$\\$L=\{(m_1,\mathsf{sensitive})\}$\\$O=\emptyset$};
\node[badnode, minimum width=3.2cm, right=2.35cm of l3] (l4) {$P=\emptyset,\ D=\{m_1,m_2\}$\\$L=\{(m_1,\mathsf{sensitive})\}$\\$O=\{\mathsf{prevent\_sensitive\_failure}\}$};

\path[trans]
    (s0) edge node[action, above] {$Q.\mathsf{submit}(m_1)$} (p1)
    (p1) edge node[action, above] {$Q.\mathsf{fail}(m_1)$} (d1)
    (d1) edge node[action, above] {$L.\mathsf{affix}(m_1,\mathsf{sensitive})$} (d1l)
    (d1l) edge node[action, above] {$Q.\mathsf{retry}(m_1)$} (repaired)
    (p1) edge node[action, left] {$L.\mathsf{affix}(m_1,\mathsf{sensitive})$} (l1)
    (l1) edge node[action, below] {$Q.\mathsf{submit}(m_2)$} (l2)
    (l2) edge node[action, below] {$Q.\mathsf{fail}(m_2)$} (l3)
    (l3) edge node[action, below] {$Q.\mathsf{fail}(m_1)$} (l4);
\end{tikzpicture}
}
\caption{Fragment of the transition system obtained by composing the \inlineconcept{DeadLetterQueue} and \inlineconcept{Label} concepts with the reactions in Listing~\ref{lst:sensitive}, with $\mathsf{Msg}=\{m_1,m_2\}$ and $\mathsf{Tag}=\{\mathsf{sensitive}\}$. States record the pending messages $P$, dead messages $D$, label relation $L$, and ongoing reaction obligations $O$, written as reaction instances. Only representative transitions are shown; highlighted states violate the no-dead-sensitive-message invariant while a reaction obligation is pending.}
\label{fig:label-dlq-product}
\end{figure}
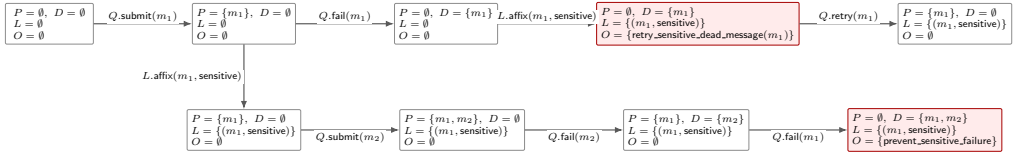

Finally, the semiautomaton of an application can be obtained by a synchronous composition of the concepts and the reaction monitor. Given $C = (S_C,\Sigma,\delta_C,i_C)$ and $\mathcal{M}_\mathcal{R} = (S_\mathcal{R},S_C \times \Sigma, \delta_\mathcal{R}, \emptyset)$ the application automaton is defined as
\begin{displaymath}
    C \otimes \mathcal{M}_\mathcal{R} = (S_C \times S_\mathcal{R}, \Sigma, \delta, (i_C, \emptyset))
\end{displaymath}
where the transition function is defined as
\[
    \delta((s,r),\sigma) =
(\delta_C(s,\sigma),\delta_\mathcal{R}(r,(s,\sigma))) 
\]
if $\delta_C(s,\sigma)$ and $\delta_\mathcal{R}(r,(s,\sigma))$ are defined. Figure~\ref{fig:label-dlq-product} shows an excerpt of the automaton for the application in Listing~\ref{lst:sensitive}.

\subsection{Verifying Invariants and Scenarios}

With the application semantics in place, verification reduces to checking temporal properties over the composed concept state and reaction-monitor state. An invariant is a first-order formula that defines a predicate over $S_C$, the global state of the concepts. Let $(s,r)$ be the current application state, where $s \in S_C$ is the current concept state and $r \in S_\mathcal{R}$ is the current reaction-monitor state. We say that the application is reacting whenever some reaction obligation is pending:
\begin{displaymath}
    \mathsf{reacting}(s,r) \iff r \neq \emptyset.
\end{displaymath}
When an invariant is broken, reactions may temporarily repair it or raise an error if the failure is fatal. As such, invariant preservation should be checked only in states where the application is not reacting. We also check that the application does not overreact, for example by raising an error when the invariant already holds. Using linear temporal logic, we can express these two safety requirements as follows:
\begin{displaymath}
    \square(\neg\mathsf{reacting}(s,r) \Rightarrow I(s))
\end{displaymath}
\begin{displaymath}
    \square(I(s) \Rightarrow \neg\mathsf{reacting}(s,r)).
\end{displaymath}
Together, these requirements state that an application is settled exactly in states where the invariant holds. For example, the design in Listing~\ref{lst:sensitive} satisfies both these properties. In the automaton excerpt in Figure~\ref{fig:label-dlq-product}, this means that the highlighted states are precisely the states with pending obligations: they may violate the invariant, but only while the application is reacting.

The semantics can be encoded directly in high-level state-based formal methods such as Alloy 6~\cite{jackson2012software,macedo2016lightweight} or TLA+~\cite{lamport1994temporal}, and the corresponding model checkers can then verify these requirements. To encode asynchronous concept composition, we include an action variable $a$ that nondeterministically chooses the next action to execute. These (and most) model checkers require the transition relation to be total. We ensure that by making the system stutter when the transition function is not defined. In \tool we use Alloy 6 because it supports full linear temporal logic, including the next-state operator $\bigcirc$, which also lets us verify whether particular scenarios are possible.

Given a scenario $A_0; \ldots; A_n$, we check whether there exists a behavior whose action prefix is the scenario and that can subsequently remain settled forever. This is done by asking the model checker to find a counterexample to the following formula:
\begin{displaymath}
\lozenge\square\neg\mathsf{reacting}(s,r)
\Rightarrow
\neg\left(\bigwedge_{i=0}^{n} \bigcirc^i(a = A_i)\right)
\end{displaymath}
If this property has a counterexample, then the scenario is possible. If it has no counterexample, then no behavior with the given action prefix can be extended to a behavior that eventually remains settled.

\section{LLM-Driven Synthesis}
\label{sec:synthesis}

The semantics above make invariants and scenarios mechanically checkable, which in turn allows them to be used as feedback in a synthesis loop. We now describe the LLM-driven synthesis procedure implemented in \tool. Rather than relying on a symbolic synthesizer, \tool asks an LLM to propose candidate reactions and uses bounded model checking to reject candidates that violate the invariant or the classified scenarios. We also describe an LLM-driven scenario elicitation procedure that helps users explore the design space by classifying proposed scenarios instead of authoring all of them from scratch. Both techniques are implemented in the open-source \tool prototype.

\subsection{Scenario-Guided Synthesis}

\begin{figure}[t]
\centering
\resizebox{\linewidth}{!}{%
\begin{tikzpicture}[
  font=\scriptsize,
  >=Stealth,
  node distance=4mm and 11mm,
  box/.style={
    draw=black!55,
    rounded corners=1pt,
    align=center,
    inner sep=4pt,
    minimum height=8mm,
    text width=2.65cm
  },
  data/.style={box, fill=gray!8},
  llm/.style={box, fill=blue!6},
  check/.style={box, fill=orange!10},
  fail/.style={box, fill=red!6},
  ok/.style={box, fill=green!10},
  flow/.style={->, thick, black!70},
  lab/.style={font=\scriptsize, fill=white, inner sep=1pt}
]
\node[data] (input) {Concepts + partial app with invariant and (optional) scenarios};
\node[llm, below=of input] (react) {Propose new or revise existing reactions};
\node[data, below=of react] (app) {Candidate complete app};
\node[llm, below=of app] (alloy) {Translate to Alloy};
\node[check, below=of alloy] (verify) {Verify all scenarios and invariant};
\node[ok, below=of verify] (done) {Verified reaction design};

\node[fail, left=of verify] (feedback) {Combined feedback with scenario mismatches and counterexample trace};
\node[fail, right=of verify] (repair) {Alloy execution log};

\draw[flow] (input) -- (react);
\draw[flow] (react) -- (app);
\draw[flow] (app) -- (alloy);
\draw[flow] (alloy) -- (verify);
\draw[flow] (verify) -- node[lab, right] {all pass} (done);

\draw[flow] (verify.west) -- node[lab, above] {issue} (feedback.east);
\draw[flow] (feedback.north) |- node[lab, near end, above] {revise} (react.west);

\draw[flow] (verify.east) -- node[lab, above] {error} (repair.west);
\draw[flow] (repair.north) |- node[lab, near end, above] {retry} (alloy.east);
\end{tikzpicture}
}
\caption{Scenario-guided synthesis loop implemented by \tool. The reaction model completes the partial app with reactions. Each candidate app is translated to Alloy and then verified against both the classified scenarios and the desired invariant. Scenario mismatches and invariant counterexamples are collected as feedback for the next reaction-model revision; Alloy translation errors are repaired separately.}
\label{fig:scenario-guided-synthesis}
\end{figure}
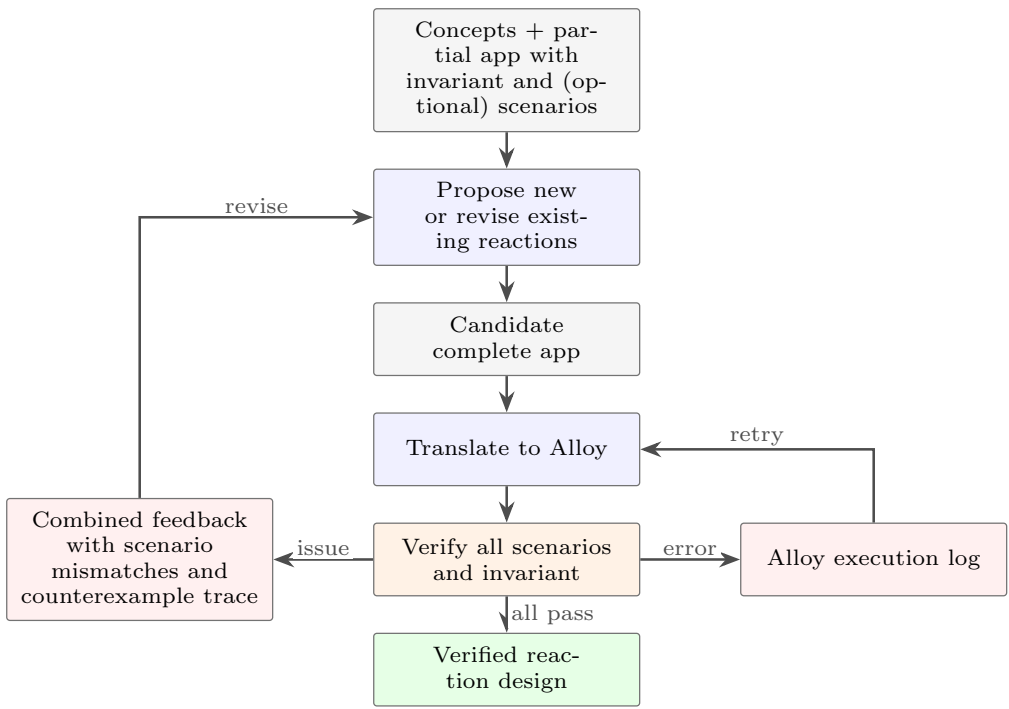

The CEGIS-style synthesis loop implemented in \tool is depicted in Figure~\ref{fig:scenario-guided-synthesis}. Two different agents (depicted as blue boxes) are used: one to propose or revise reactions that satisfy the desired invariant and scenarios, and another to translate the invariant, scenarios, reactions and concept markdown descriptions to Alloy, encoding the semantics defined in the previous section. This second agent is not essential to our technique, nor a claimed contribution of the paper, but is needed at the moment because we currently use logic conditions expressed using natural-language, which are not amenable for automatic translation. In the future we intend to define a formal language that enables that automatic translation, but still retains a natural-language flavor that is LLM-friendly.

The verification is done with the Alloy 6 Analyzer, using the bounded model checking engine. A scenario mismatch means that the Alloy analysis disagrees with the user classification: an \texttt{ok} scenario is not possible, or a \texttt{nok} scenario is possible. In each synthesis iteration, \tool checks all scenario commands and the desired invariant. All observed scenario mismatches and any invariant counterexample found by the Analyzer are collected and sent as feedback for the reaction agent to propose new reactions or revise existing ones. The tool also includes a feedback loop if there are Alloy execution errors (typically syntax errors), giving the Alloy translation agent the opportunity to fix those. If there are no scenario mismatches or invariant counterexamples, the design is considered verified within the chosen Alloy scope and step bound and is output. If the maximum number of synthesis iterations is reached first, synthesis reports failure and preserves the last generated candidate and verification artifacts for inspection.

The system and task prompts given to the agents are rather standard. The system prompt includes a brief overview of Concept Design and a few examples of concepts and reactions (using concepts different from those in our evaluation benchmark). The reaction task instructs the model to use only the provided concepts and actions, to leave the invariant and scenarios unchanged, and to make each reaction trigger an action in one concept or the special error action. In repair iterations, the prompt includes the accumulated verification feedback so that the model does not fix the latest problem by reintroducing an earlier one. The Alloy translation prompt includes a couple of examples of how concepts should be translated to Alloy, and one example of translating one complete application with reactions and scenarios.

\subsection{Scenario Elicitation}

\begin{figure}[t]
\centering
\resizebox{\linewidth}{!}{%
\begin{tikzpicture}[
  font=\scriptsize,
  >=Stealth,
  node distance=5mm and 10mm,
  box/.style={
    draw=black!55,
    rounded corners=1pt,
    align=center,
    inner sep=4pt,
    minimum height=8mm,
    text width=2.65cm
  },
  data/.style={box, fill=gray!8},
  model/.style={box, fill=blue!6},
  ask/.style={box, fill=orange!10},
  nok/.style={box, fill=green!10},
  ok/.style={box, fill=green!10},
  flow/.style={->, thick, black!70},
  lab/.style={font=\scriptsize, fill=white, inner sep=1pt}
]
\node[data] (context) {Concepts + partial app with invariant and (optional) scenarios};
\node[model, below=of context] (prefix) {Propose short\\invariant-breaking\\prefix $P$};
\node[ask, below=of prefix] (classp) {User classifies\\$P$};

\node[nok, left=of classp] (rejectp) {Record \texttt{nok}: $P$\\close $P$};
\node[ok, right=of classp] (acceptp) {Record \texttt{ok}: $P$\\keep $P$ open};
\node[model, below=of acceptp] (repair) {Propose continuation\\$P;F$ with candidate\\repair suffix $F$};
\node[ask, below=of repair] (classf) {User classifies\\$P;F$};
\node[nok, left=of classf] (rejectf) {Record \texttt{nok}: $P;F$\\close $P;F$};
\node[ok, right=of classf] (acceptf) {Record \texttt{ok}: $P;F$\\close $P$};

\draw[flow] (context) -- (prefix);
\draw[flow] (prefix) -- (classp);
\draw[flow] (classp) -- node[lab, above] {\texttt{nok}} (rejectp);
\draw[flow] (classp) -- node[lab, above] {\texttt{ok}} (acceptp);
\draw[flow] (acceptp) -- (repair);
\draw[flow] (repair) -- (classf);
\draw[flow] (classf) -- node[lab, above] {\texttt{nok}} (rejectf);
\draw[flow] (classf) -- node[lab, above] {\texttt{ok}} (acceptf);

\draw[flow] (rejectf.north) |- node[lab, near end, left] {vary $F$} (repair.west);
\draw[flow] (rejectp.north) |- node[lab, near end, above] {vary $P$} (prefix.west);
\draw[flow] (acceptf.north) |- node[lab, near end, above] {vary $P$} (prefix.east);
\end{tikzpicture}
}
\caption{Scenario elicitation strategy used by \tool. The model first asks about short prefixes $P$ that make the invariant invalid. Each user classification is recorded as a scenario. If $P$ is classified \texttt{nok}, the model closes that prefix and varies $P$. If $P$ is classified \texttt{ok}, the model keeps that prefix open and asks about continuations $P;F$, closing rejected continuations and varying $F$ until a repair suffix is accepted.}
\label{fig:scenario-elicitation}
\end{figure}
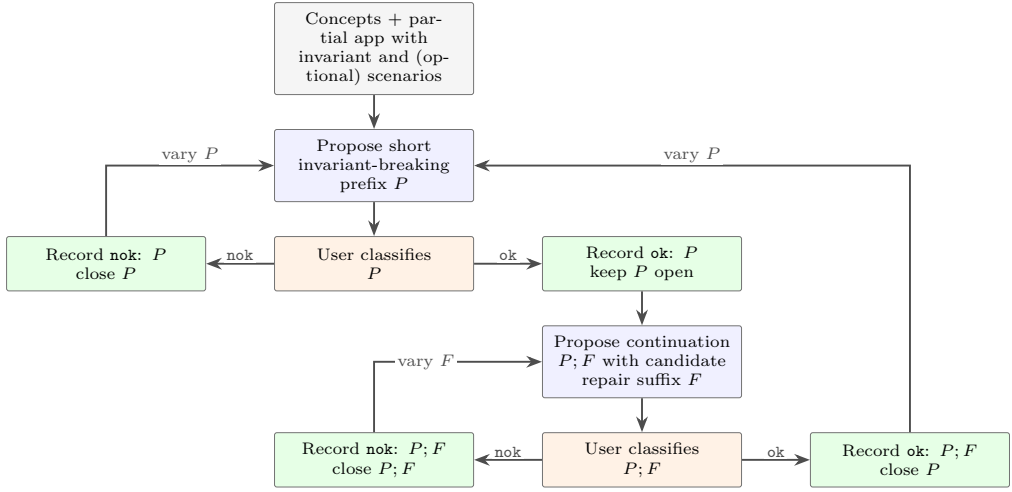

The scenario elicitation strategy used by \tool is described in Figure~\ref{fig:scenario-elicitation}. Given a partial application description with the desired invariant and, optionally, some scenarios already classified, the LLM is first asked to propose a single unclassified scenario: a short prefix of actions $P$ that leads to an invariant violation and is semantically different from the scenarios already explored. Semantic difference is defined by the final action and the observable pre-state of that action, since reactions can only depend on the current action and state, not on the history that produced that state. The user is then asked to classify this prefix as a desirable (\texttt{ok}) or undesirable (\texttt{nok}) sequence of actions. If the user says it is undesirable, the prefix $P$ is added to the current set of negative scenarios, and the LLM is asked to vary $P$. If the prefix is classified as desirable, it is added as a positive scenario, but the LLM is then asked to propose candidate suffixes $F$ that could repair the invariant violation. For every rejected suffix $F$, $P;F$ is added as a negative scenario. Once a suffix $F$ is accepted by the user, $P;F$ is added as a positive scenario, and the model is instructed to explore different invariant-violating prefixes.

\section{Evaluation}
\label{sec:evaluation}

Our evaluation of \tool aims to answer the following research questions:
\begin{description}
    \item[RQ1] How is the efficiency of design verification with Alloy 6 affected by the domain scope and step bound?
    \item[RQ2] Given only the invariant, how often is counterexample-guided refinement needed to reach a verified design, and how stable and plausible are the resulting designs?
    \item[RQ3] How does scenario-guided synthesis compare with prompt-guided synthesis in steering the result toward the intended verified design?
    \item[RQ4] Can LLM-driven scenario elicitation reduce user specification effort while preserving enough design coverage to recover the intended design?
    \item[RQ5] What failure modes typically arise in scenario-guided synthesis?
\end{description}
We focus the evaluation on one GPT LLM configuration: \texttt{gpt-5.5}, used with the default parameters, namely temperature 1 and medium reasoning effort. All experiments were run in June 2026, and the \tool tool and all synthesized designs are publicly available at \url{https://github.com/alcinocunha/foundry}.

\subsection{Benchmark}

To answer these research questions, we need a benchmark of applications, each with multiple alternative designs. We began by defining three applications with critical safety invariants. Table~\ref{tab:benchmark} summarizes these applications and their respective invariants. 

\begin{table}
    \centering
  \caption{Application benchmark}
  \label{tab:benchmark}
  \begin{tabular}{lp{0.15\linewidth}p{0.5\linewidth}}
    \toprule
    Application & Concepts & Invariants \\
    \midrule
    \inlineconcept{NoDeadSensitiveMessages} & \inlineconcept{DeadLetterQueue}\inlineconcept{Label} & There are no sensitive messages in the dead letter queue. \\
    \inlineconcept{CourseManagementSystem} & \inlineconcept{Rostering} \inlineconcept{Registration} & Users enrolled in the course (aka students) must be registered in the forum with permission role Guest. If there is someone registered in the forum, then at least one user must be registered as Admin. \\
    \inlineconcept{OneTimeEvidenceLinks} & \inlineconcept{Vault} \inlineconcept{Permalink} & Evidence files with non-revoked permalinks must be decrypted. Decrypted evidence files must have exactly one non-revoked permalink. Accessed permalinks must be revoked. \\
    \bottomrule
\end{tabular}
\end{table}

The first is the same application used in the overview, composing the \inlineconcept{DeadLetterQueue} and \inlineconcept{Label} concepts. A special ``sensitive'' label marks messages that contain sensitive data, and the invariant states that such messages cannot remain in the \emph{dead-letter queue}, where they risk being leaked to less protected operational storage.

The second application, \inlineconcept{CourseManagementSystem}, composes a \inlineconcept{Rostering} concept, which tracks the students enrolled in a course, with a \inlineconcept{Registration} concept, which records the users registered in the course's public forum and manages the corresponding permission roles. The invariants ensure that enrolled students are registered in the forum with the Guest permission role and that, whenever someone is registered in the forum, at least one user is registered as Admin.

The third application, \inlineconcept{OneTimeEvidenceLinks}, composes the \inlineconcept{Vault} and \inlineconcept{Permalink} concepts. The former provides actions to create, decrypt, and encrypt data, with newly created data encrypted by default; the latter supports sharing items through permalinks, which can later be accessed or revoked. The application supports temporary disclosure of evidence files through single-use sharing permalinks. Evidence files are kept encrypted in a vault and are decrypted only temporarily for sharing. The invariants ensure that the only unencrypted evidence files are those currently being shared and that permalinks are single-use.

\begin{table}
    \centering
  \caption{Benchmark of design variants. The intended behavior column summarizes the distinguishing reactions of each variant; problematic behaviors not described there are forbidden and raise an error.}
  \label{tab:variants}
  \begin{tabular}{c|p{0.7\linewidth}|c}
    \toprule
    Variant & \multicolumn{1}{|c|}{Intended behavior} & Reactions \\
    \midrule
    \multicolumn{3}{c}{\inlineconcept{NoDeadSensitiveMessages}}\\
    \midrule
    1 & Tagging a dead message as sensitive triggers a retry. & 2\\
    2 & Tagging a dead message as sensitive triggers a purge. & 2 \\
    3 & Tagging a dead message as sensitive triggers a purge if there are no other dead messages, otherwise triggers a retry. & 3 \\
    4 & Tagging a dead message as sensitive triggers a retry of other dead messages and then a purge of the sensitive message. & 4 \\
    \midrule
    \multicolumn{3}{c}{\inlineconcept{CourseManagementSystem}}\\
    \midrule
    1 & The enrollment of a non registered user triggers their registration if the forum has some Admin. & 7 \\
    2 & The enrollment of a non registered user triggers their registration. If a forum Admin enrolls and there is another Admin, permissions are changed to Guest. If a user enrolls and there is no Admin or if the only Admin enrolls a new admin is first registered and then the user is registered. & 11 \\
    3 & The enrollment of a non registered user triggers their registration if the forum has some Admin. Last admin can deregister if there are no forum members also enrolled in the course, causing all forum members to be deregistered. & 8 \\
    4 & The enrollment of a non registered user triggers their registration if the forum has some Admin. Last admin can deregister or be assigned the Guest role if there is another forum member that can assume the Admin role. & 9 \\
    \midrule
    \multicolumn{3}{c}{\inlineconcept{OneTimeEvidenceLinks}}\\
    \midrule
    1 & Sharing an encrypted file triggers decrypting. Access triggers revoking and encrypting in any order. & 8 \\
    2 & Sharing an encrypted file triggers decrypting and decrypting triggers sharing. Access triggers revoking and then encrypting. & 8 \\
    3 & Sharing an encrypted file triggers decrypting. Access triggers revoking and then resharing. &  8 \\
    4 & Sharing one encrypted file triggers the sharing of all encrypted files and only then they are decrypted (in any order). Access triggers revoking and then encrypting. & 9 \\
    \bottomrule
\end{tabular}
\end{table}

For each application, we then defined multiple plausible designs that satisfy the invariant. Table~\ref{tab:variants} presents all the design variants in our benchmark, namely how the application should react to some problematic behaviors. If a problematic situation is not described in the table, that situation is forbidden and raises an error. The table also reports how many reactions were manually specified by the author for each design. 

It is worth noting that the verification mode of \tool was essential in this process, as the author often made subtle mistakes in the manual encoding of the reactions, even after specifying the desired scenarios first. The only variants for which the author specified the reactions correctly on the first attempt were the first three variants of the \inlineconcept{NoDeadSensitiveMessages} application. This shows that, even for such relatively simple designs, formal verification techniques are essential to ensure correctness. Unlike the author, for some of these variants the LLM actually produced a correct design on the first attempt, as discussed below.

\subsection{Efficiency of Design Verification}

To answer RQ1, we formally verified the manually specified designs for all application variants using Alloy Analyzer 6.3 with nuXmv 2.1 as the bounded model-checking backend. For each variant, we measured verification time while varying the domain scope from 2 to 4 with a fixed bound of 10 steps, and while varying the bound from 8 to 12 steps with a fixed scope of 3. The domain scope sets the maximum number of entities in each type, while the step bound limits the length of the traces that are analyzed by the bounded model checking procedure. Table~\ref{tab:verification_times} reports average running times in seconds (over three runs), measured on a commodity laptop with an Apple M1 processor and 16 GB of memory.

\begin{table}
    \caption{Design verification times, in seconds. The left block varies the scope with a fixed bound of 10 steps; the right block varies the bound with a fixed scope of 3. Bold entries correspond to the default setting used in the remaining experiments.}
    \label{tab:verification_times}
    \centering
    \small
    \setlength{\tabcolsep}{4pt}
    \begin{tabular}{lrrrrrrrr}
        \toprule
        & \multicolumn{3}{c}{Scope, 10 steps}
        & \multicolumn{5}{c}{Steps, scope 3} \\
        \cmidrule(lr){2-4}
        \cmidrule(lr){5-9}
        Variant & 2 & 3 & 4 & 8 & 9 & 10 & 11 & 12 \\
        \midrule
        \multicolumn{9}{c}{\inlineconcept{NoDeadSensitiveMessages}}\\
        \midrule
        1 & 1.16 & \textbf{2.16} & 3.73 & 1.42 & 1.76 & \textbf{2.16} & 3.29 & 2.92 \\
        2 & 1.00 & \textbf{1.37} & 2.35 & 1.03 & 1.22 & \textbf{1.37} & 1.59 & 1.84 \\
        3 & 1.18 & \textbf{2.31} & 3.98 & 1.37 & 1.71 & \textbf{2.31} & 2.51 & 3.81 \\
        4 & 1.52 & \textbf{2.34} & 6.14 & 1.56 & 1.81 & \textbf{2.34} & 3.15 & 4.74 \\
        \midrule
        \multicolumn{9}{c}{\inlineconcept{CourseManagementSystem}}\\
        \midrule
        1 & 1.70 & \textbf{11.21} & 90.82 & 3.52 & 7.56 & \textbf{11.21} & 17.65 & 32.77 \\
        2 & 2.73 & \textbf{43.85} & 593.64 & 9.57 & 15.23 & \textbf{43.85} & 63.85 & 156.04 \\
        3 & 2.02 & \textbf{21.04} & 134.16 & 4.20 & 7.88 & \textbf{21.04} & 44.40 & 76.27 \\
        4 & 3.47 & \textbf{34.81} & 745.48 & 6.00 & 15.00 & \textbf{34.81} & 74.35 & 233.24 \\
        \midrule
        \multicolumn{9}{c}{\inlineconcept{OneTimeEvidenceLinks}}\\
        \midrule
        1 & 2.76 & \textbf{21.00} & 88.52 & 4.52 & 8.92 & \textbf{21.00} & 49.73 & 135.36 \\
        2 & 3.02 & \textbf{42.88} & 164.29 & 6.67 & 24.46 & \textbf{42.88} & 117.24 & 256.69 \\
        3 & 2.57 & \textbf{15.76} & 47.83 & 4.34 & 8.67 & \textbf{15.76} & 34.52 & 89.53 \\
        4 & 3.07 & \textbf{31.18} & 129.91 & 6.35 & 15.27 & \textbf{31.18} & 63.50 & 158.54 \\
        \bottomrule
    \end{tabular}
\end{table}

As expected, increasing the scope has a far steeper impact on verification time than increasing the number of bounded steps. Each unit increase in scope multiplies verification time by a large factor---at least an order of magnitude for some of the \inlineconcept{CourseManagementSystem} variants---consistent with the exponential blowup expected from bounded analysis, whereas increasing the number of steps tends to raise times much more gently. 

A scope of 3 with 10 steps (incidentally, the defaults used by the Alloy Analyzer) seems to provide a good tradeoff between assurance and efficiency, with verification times under one minute for all variants. This matters because synthesis invokes the verifier once per iteration: keeping each verification call under a minute is what makes the counterexample-guided loop practical at these bounds. Those were therefore the bounds set for the remaining experiments in our evaluation. For very critical designs they could be pushed higher, but synthesis would likely need to be run offline, owing to the longer total synthesis time.

Overall, for this benchmark, verification is fast enough to support interactive synthesis at scope 3 and 10 steps, and scalability is mainly limited by domain scope.

\subsection{Invariant-guided Synthesis}

To answer RQ2, we ran our synthesis procedure three times for each application, providing only the invariant as input.  The number of iterations required to converge to a verified design in each run is reported in Table~\ref{tab:invariant-synthesis-results}. The maximum number of iterations was set to 10.

\begin{table}
    \centering
    \caption{Invariant-guided synthesis results. Each iteration entry contains three independent runs.}
    \label{tab:invariant-synthesis-results}
    \begin{tabular}{l|c|c|c}
        \toprule
        Application & Iterations & Distinct designs & Plausible designs \\
        \midrule
        \inlineconcept{NoDeadSensitiveMessages} & 1 1 1 & 2 & 0/3 \\
        \inlineconcept{CourseManagementSystem} & 2 1 1 & 3 & 3/3 \\
        \inlineconcept{OneTimeEvidenceLinks} & 2 1 1 & 3 & 3/3 \\ 
        \bottomrule
    \end{tabular}
\end{table}

Across the nine runs, GPT-5.5 produced a verified design on the first attempt in seven runs, but in two runs the initial design violated the invariant and required one counterexample-guided refinement step. This shows that even in the invariant-only setting, where the goal is simply to find any invariant-satisfying design, LLM-generated reactions cannot be trusted without formal verification: the verifier is needed both to detect incorrect designs and to provide counterexamples that guide their repair.

Reaching a verified design, however, does not mean that the design is stable across runs or plausible for the application domain. Therefore, we also manually inspected the designs to assess whether they produced the same design and the plausibility of the resulting designs. A design was deemed implausible if it failed an implicit, widely-understood expectation in the application's domain that cannot be expressed as an invariant (at least in our current formalization).

In the \inlineconcept{NoDeadSensitiveMessages} application we obtained two different designs, both of which we deemed implausible. In one run, the synthesized design simply forbade both problematic behaviors: failing a sensitive message and classifying an already dead message as sensitive. Although this preserves the invariant, it is not a reasonable design because the message is already sensitive; forbidding the label merely prevents the system from recording that fact, allowing the sensitive data to remain in the dead-letter queue without being treated as sensitive. In the other two runs, the synthesized design repaired invariant violations by detaching the sensitive label. This is also implausible, because it enforces the invariant by declassifying sensitive data rather than by moving or deleting the message. For the other two applications, invariant-only synthesis produced three distinct but plausible designs in each case, confirming that the invariant alone often leaves several reasonable coordination policies open.

These results support our expectation that invariants alone provide too much freedom when exploring the design space. Although all runs reached a verified design within two iterations, only one application produced repeated designs; two applications produced three distinct designs, and one application produced only implausible designs. Additional specification artifacts, ideally with a precise semantics that can be automatically verified, are therefore needed to steer synthesis toward the user's intended design. Our proposal in this paper is to complement invariants with scenarios.

\subsection{Scenario- vs Prompt-guided Synthesis}

To assess the effectiveness of scenarios in guiding the synthesis towards the intended design, we need a baseline against which to compare. We opted for a comparison with the straightforward way of steering LLMs, namely natural-language prompts. To do that we modified \tool to also allow a natural-language prompt to be given as input together with the desired invariant.

The methodology to write the prompts was the following: we started with the description in Table~\ref{tab:variants} and added a sentence ``All other problematic behaviors must be forbidden''. If the synthesis with this initial prompt produced an unintended design, we did one revision to attempt to clarify the divergent behaviors. 

The methodology to specify the scenarios was the following. 
For each problematic behavior we defined a minimal prefix of actions that led up to the state where the invariant was broken. If that behavior is forbidden, the scenario was marked as \emph{nok}. If the behavior is allowed, we appended the intended repairing action and marked the scenario as \emph{ok}. If repairing took more than one action, we added positive scenarios for all possible repairs and negative scenarios for the disallowed ones. As with prompts, if the synthesis with this initial (minimal) set of scenarios produced an unintended design, we did one revision adding a few more scenarios to attempt to clarify the divergent behaviors. 

For example, for the third design variant of application \inlineconcept{NoDeadSensitiveMessages} in Table~\ref{tab:variants}, the initial set of scenarios was
\begin{lstlisting}[style=pacmplstyle, numbers=none, language=concepts]
nok: Q.submit(m); L.affix(m,Sensitive); Q.fail(m);
ok:  Q.submit(m); Q.fail(m); L.affix(m,Sensitive); Q.purge();
ok:  Q.submit(m); Q.fail(m); Q.submit(n); Q.fail(n); 
     L.affix(n,Sensitive); Q.retry(n);
\end{lstlisting}
whereas, for the fourth variant, the last scenario was replaced by the following.
\begin{lstlisting}[style=pacmplstyle, numbers=none, language=concepts]
ok:  Q.submit(m); Q.fail(m); Q.submit(n); Q.fail(n); 
     L.affix(n,Sensitive); Q.retry(m); Q.purge();
\end{lstlisting}

We ran the synthesis procedure three times and manually checked whether the synthesized designs were equivalent to the intended one. To increase our confidence in this analysis, for the designs synthesized with the prompts we also checked whether they conform to the scenarios used for the scenario-guided synthesis. Table~\ref{tab:synthesis-results} reports the results of the experiment, namely for each variant the number of iterations GPT-5.5 needed to converge to a verified design in three different synthesis attempts. The maximum number of iterations was set to 10. For the scenario-guided synthesis, we also list the number of positive and negative scenarios that were specified. We report in red the cases where the resulting design was not equivalent to the intended one. As mentioned above, when this happened at least once with the initial prompt / scenarios, we revised them and repeated the experiment. 

\begin{table}
    \centering
  \caption{Prompt and scenario-guided synthesis results. Each iteration entry contains three independent runs; empty revised columns mean no revision was needed.}
  \label{tab:synthesis-results} 
  \begin{tabular}{c|c|c|c@{ }c|c|c@{ }c|c}
    \toprule
    & \multicolumn{2}{|c|}{Prompt-guided} & \multicolumn{6}{|c}{Scenario-guided} \\
    Variant & Initial & Revised & Pos & Neg & Initial & Pos & Neg & Revised \\
    \midrule
    \multicolumn{9}{c}{\inlineconcept{NoDeadSensitiveMessages}}\\
    \midrule
     1 & 1 1 1 & & 1 & 1 & 1 1 1 & & & \\
     2 & 1 1 1 & & 1 & 1 & 1 1 1 & & & \\
     3 & 1 1 1 & & 2 & 1 & 2 1 2 & & & \\ 
     4 & 1 1 2 & & 2 & 1 & \textcolor{red}{3} \textcolor{red}{6} \textcolor{red}{4} & 3 & 1 & 1 1 5\\
    \toprule
    \multicolumn{9}{c}{\inlineconcept{CourseManagementSystem}}\\
    \midrule
     1 & 1 1 3 & & 1 & 7 & 1 1 1 & & & \\
     2 & 2 2 1 & & 4 & 7 & 2 1 6 & & & \\
     3 & 1 1 1 & & 2 & 7 & 2 2 3 & & & \\ 
     4 & \textcolor{red}{5} 6 2 & 4 3 4 & 3 & 6 & \textcolor{red}{3} 2 \textcolor{red}{3} & 7 & 6 & 2 3 3 \\
    \toprule
    \multicolumn{9}{c}{\inlineconcept{OneTimeEvidenceLinks}}\\
    \midrule
     1 & \textcolor{red}{2} 2 3 & 1 1 1 & 3 & 5 & 1 2 1 & & & \\
     2 & 2 2 \textcolor{red}{1} & 3 3 1 & 3 & 5 & 1 1 2 & & & \\
     3 & 1 1 2 & & 2 & 6 & 3 1 4 & & & \\ 
     4 & \textcolor{red}{1} \textcolor{red}{2} \textcolor{red}{2} & 1 1 \textcolor{red}{3} & 4 & 7 & 4 6 7 & & & \\
    \bottomrule
\end{tabular}
\end{table}

In the last variant of \inlineconcept{NoDeadSensitiveMessages}, all runs of the scenario-guided synthesis failed to produce the intended design. This was due to an overfitting failure, where the model failed to generalize from the given scenarios: when a dead message was marked as sensitive it only retried the non-sensitive messages when there were exactly two dead messages. This suggests that providing only a minimal set of scenarios may not be ideal, especially for cascading reactions. These overfitting failures will also be discussed in Section~\ref{sec:failure_modes}. In the revised scenarios we added the following positive scenario, which was sufficient to make the synthesis converge to the intended design in all runs.
\begin{lstlisting}[style=pacmplstyle, numbers=none, language=concepts]
ok: Q.submit(m); Q.fail(m); Q.submit(n); Q.fail(n); Q.submit(s); 
    Q.fail(s); L.affix(s,Sensitive); Q.retry(n); Q.retry(m); Q.purge();
\end{lstlisting}
The prompt-guided synthesis performed very well in this application, almost always producing the intended design on the first attempt.

A similar problem occurred in the last variant of application \inlineconcept{CourseManagementSystem} with scenario-guided synthesis: the synthesized design only allowed the deregistration or role change of the last Admin if exactly one other user was registered as Guest, because those were the minimal scenarios that were provided. In the revised scenarios we added extra cases with two registered Guest users, and the synthesis always converged to the intended design. A similar overfitting problem also occurred in one of the prompt-guided runs. To try to avoid that issue we added the following sentence to the initial prompt: ``In these situations if there is more than one forum Guest member that can assume the Admin role, one of them must be promoted to Admin''. With this revised prompt, the synthesis always converged to the intended design.

In \inlineconcept{OneTimeEvidenceLinks} we had several issues with the prompt-guided synthesis. A recurrent issue, that occurred at least once in all but the third variant, was that the synthesized design allowed revoking shared permalinks that were not accessed. In the revised prompts we added the sentence ``Revoking a non accessed url must be forbidden'' to reinforce that this is not allowed. That was sufficient to solve the issue in the revised runs. 
In all the runs of the last variant the synthesized design also committed on a specific order of decrypting shared files, namely the first file to be decrypted had to be the last to be shared. To address this issue we replaced the ``in any order'' by a more explicit ``in this situation the system should be allowed to decrypt the files in any order''. 

Unfortunately in one of the runs with the revised prompt the design still committed on a specific order. 
The remaining revised-prompt failure was not merely an ordering artifact: the design had an error that prevented sharing the last file. The problematic reaction was the following.
\begin{lstlisting}[style=pacmplstyle, numbers=none, language=concepts]
reaction share_unshared_encrypted_file_before_decryption
when  P.share(e,u)
where e is in encrypted in V
      x is in encrypted in V
      x has no non revoked url in P
then  some t : Token | P.share(x,t)
\end{lstlisting}
The \emph{where} clause does not require \inlineconcept{e} and \inlineconcept{x} to be different, so sharing the last file would require sharing it again, which in turn causes an error because it is not possible to have more than one non-revoked url. Raising an error in this situation is a possible way to avoid breaking the invariant, but was not the intended design. 

Overall, scenario-guided synthesis fared well when compared to prompt-guided synthesis. A minimal set of scenarios sufficed to consistently produce the intended design in 10 out of 12 variants, while our first prompt attempt only achieved that on 8 out of 12 variants. The revised scenarios fixed both initial failures; prompts still failed in one variant after revision. In both techniques the counterexample-guided synthesis consistently helped the LLM reach verified designs, with an average of 1.79 iterations for prompt-guided runs, 2.36 iterations for scenario-guided runs, and 2.06 iterations across all 90 runs.

As we have shown, scenario-guided synthesis is prone to overfitting problems if only minimal scenarios are provided. This suggests a practical mitigation: include scenarios that vary cardinalities in cascading or quantified repairs. A clear advantage of prompts is that they can be very economical when compared to scenarios, because natural-language can quantify informally. For example, the single sentence at the end of our prompts --- ``All other problematic behaviors must be forbidden'' --- has no counterpart in scenarios, and we need to enumerate all forbidden problematic behaviors. On the other hand, prompts are very brittle. Often the LLM fails to follow instructions that seem quite clear, and it may not be easy to find a wording that clarifies them. In our experience, scenarios were easier to revise, because revision is purely additive and rather mechanical. More importantly, prompts, unlike scenarios, are not verifiable, and thus are not a good specification artifact to steer automatic synthesis procedures.

\subsection{Scenario Elicitation}

To answer RQ4, we used GPT-5.5 to perform scenario elicitation twice for all variants: one run elicited 10 scenarios and the other 20 scenarios. We then ran the scenario-guided synthesis procedure three times for each batch of classified scenarios. 
Table~\ref{tab:scenario-elicitation-results} presents the results, namely the number of positive and negative scenarios classified in each batch, and the number of iterations GPT-5.5 then took to synthesize a verified design. Again, we report in red the cases where the resulting design was not equivalent to the intended one. A red dash means that GPT-5.5 refused to produce a design at some point before reaching the limit of 10 iterations because it believed the scenarios were inconsistent. Besides the manual inspection, we also used the scenarios defined manually for the previous RQ to further check whether the produced design was equivalent.

\begin{table}
    \centering
  \caption{Scenario elicitation results}
  \label{tab:scenario-elicitation-results}
  \begin{tabular}{c|c@{ }c|c|c@{ }c|c|c}
    \toprule
    & \multicolumn{3}{|c|}{10 Scenarios} & \multicolumn{3}{|c|}{20 Scenarios} & \\
    Variant & Pos & Neg & Iterations & Pos & Neg & Iterations & Overlap \\
    \midrule
    \multicolumn{8}{c}{\inlineconcept{NoDeadSensitiveMessages}}\\
    \midrule
    1 & 8 & 2 & 1 2 3 & 12 & 8 & 1 1 1 & 5 \\
    2 & 6 & 4 & 1 1 2 & 12 & 8 & 1 2 1 & 7 \\
    3 & 5 & 5 & \textcolor{red}{--} \textcolor{red}{4} 3 & 8 & 12 & 1 1 2 & 6 \\ 
    4 & 7 & 3 & \textcolor{red}{1} \textcolor{red}{2} \textcolor{red}{2} & 8 & 12 & 2 2 1 & 6 \\
    \toprule
    \multicolumn{8}{c}{\inlineconcept{CourseManagementSystem}}\\
    \midrule
    1 & 2 & 8 & 2 \textcolor{red}{1} \textcolor{red}{1} & 6 & 14 & 3 3 4 & 8 \\
    2 & 6 & 4 & 2 5 3 & 9 & 11 & 4 2 2 & 7 \\
    3 & 4 & 6 & \textcolor{red}{1} \textcolor{red}{2} \textcolor{red}{2} & 6 & 14 & 3 5 2 & 9 \\
    4 & 6 & 4 & \textcolor{red}{3} \textcolor{red}{2} \textcolor{red}{3} & 7 & 13 & \textcolor{red}{2} \textcolor{red}{7} \textcolor{red}{4} & 6 \\
    \toprule
    \multicolumn{8}{c}{\inlineconcept{OneTimeEvidenceLinks}}\\
    \midrule
    1 & 4 & 6 & \textcolor{red}{1} \textcolor{red}{2} \textcolor{red}{1} & 13 & 7 & 2 2 2 & 9 \\
    2 & 6 & 4 & 4 3 3 & 12 & 8 & \textcolor{red}{4} \textcolor{red}{2} \textcolor{red}{2} & 8 \\
    3 & 4 & 6 & 1 1 3 & 12 & 8 & 3 1 2 & 9 \\
    4 & 5 & 5 & \textcolor{red}{2} \textcolor{red}{2} \textcolor{red}{2} & 8 & 12 & \textcolor{red}{3} 3 \textcolor{red}{2} & 10 \\
    \bottomrule
\end{tabular}
\end{table}

With 10 scenarios, the intended design was only systematically synthesized in 5 of the 12 variants. This is somewhat expected, since some of the variants have several problematic behaviors, and limiting the elicitation to 10 scenarios does not give enough room to explore all of them. In some variants, namely variant 2 of \inlineconcept{CourseManagementSystem} it was actually a bit surprising that the technique was able to elicit enough scenarios to converge to the intended design. Not all problematic scenarios are actually covered (for example there was no scenario corresponding to the enrollment of a non-registered user), but the model was able to infer the intended behavior for those cases from inspecting other cases.

With 20 scenarios, the results improved substantially, and the intended design was systematically synthesized in 9 of the 12 variants. 
None of the runs in the last variant of \inlineconcept{CourseManagementSystem} produced the intended design. The same overfitting problem already discussed in the previous section also occurred here, but the key issue was that the scenario where a non-registered user enrolls was never elicited, and the model opted to forbid that behavior, which was not the intended behavior.
Interestingly, overfitting did not occur in the last variant of \inlineconcept{NoDeadSensitiveMessages}, even if the elicitation never produced a scenario with more than two dead messages.
In the second variant of \inlineconcept{OneTimeEvidenceLinks}, surprisingly the model never proposed the scenario \inlineconcept{V.store(e); P.share(e, t);}, instead opting to explore rather similar scenarios with long sequences of actions.

In the last variant of \inlineconcept{OneTimeEvidenceLinks} in two runs the synthesized design committed to a specific order of decrypting shared files: it required the last shared file to be decrypted last, because that was the repair that it suggested when two files were in the vault.
Unfortunately the current technique for scenario elicitation does not fare well when different sequences of actions can be used to repair an invariant violation, because once a repair is chosen by the user, the model moves to a different invariant-violating behavior. Of course it could be instructed to explicitly elicit other possible repairs, but that would make it focus too much on a single problematic behavior, instead of exploring a variety of different situations. 

Interestingly, the number of synthesis iterations does not seem to be affected by the number of scenarios. The average number of iterations to converge to a verified design was 2.11 with 10 scenarios and 2.36 with 20 scenarios (the same average as with the manually curated scenarios).

To assess the effect of non-determinism we counted, for each variant, how many of the 10 scenarios classified in the first batch were proposed in the first 10 scenarios in the second batch. We consider two scenarios the same if they correspond to the same invariant violation (same pre-state and same offending action), even if the order of messages in the prefix is different. The results are reported in column \emph{Overlap} of Table~\ref{tab:scenario-elicitation-results}. 

As can be seen, non-determinism has a significant impact: only once the first 10 scenarios were identical, the minimum overlap was 5, and the average 7.5. As mentioned above, in the second variant of \inlineconcept{OneTimeEvidenceLinks} the 10-scenario batch was actually more informative about the intended design than the 20-scenario batch: this means that the impact of non-determinism can make the technique non-monotonic in the sense that eliciting more scenarios is not always necessarily better.

Overall, LLM-driven scenario elicitation shifts part of the specification task from authoring scenarios to classifying proposed ones, but preserving design coverage appears to require a sufficiently large and diverse elicited set: 20 scenarios recovered the intended design in most variants, while missed behaviors and non-determinism still prevented reliable coverage in all cases.

\subsection{Failure modes}
\label{sec:failure_modes}

To answer RQ5, we manually inspected all scenario-guided results and classified the failures of synthesis to produce the intended design. We also report some quality failures in the synthesized designs that were equivalent to the intended ones. We focus on scenario-guided synthesis because that is the main focus of our technique. Correctness failures are counted over all scenario-guided synthesis runs, while quality failures are counted only for the successful runs that produced the intended design.

\paragraph{Missing critical scenarios.} The main reason why scenario-guided synthesis may fail to produce the intended design is missing critical scenarios in the input. This was the main cause of failure in 26 of the 72 RQ4 runs. For example, as mentioned above, in the second variant of \inlineconcept{OneTimeEvidenceLinks} if the positive scenario \inlineconcept{V.store(e); P.share(e, t); V.decrypt(e)} is not included together with \inlineconcept{V.store(e); V.decrypt(e); P.share(e, t);} the model will have no way to know that the intention is to initiate the disclosure of files either by sharing or by decrypting.

\paragraph{Scenario overfitting.} The second most frequent reason to fail to produce the intended design is overfitting, the inability to generalize the intended behavior from the given scenarios. This happened 8 times across all scenario-guided runs. For example, in the fourth variant of \inlineconcept{NoDeadSensitiveMessages}, given the scenario
\begin{lstlisting}[style=pacmplstyle, numbers=none, language=concepts]
ok:  Q.submit(m); Q.fail(m); Q.submit(n); Q.fail(n); 
     L.affix(n,Sensitive); Q.retry(m); Q.purge();
\end{lstlisting}
the model opted to retry the non-sensitive messages only when exactly two existed.
\begin{lstlisting}[style=pacmplstyle, numbers=none, language=concepts]
reaction retry_other_when_exactly_two_dead_messages
when  L.affix(n,Sensitive)
where n is in dead in Q
      m is in dead in Q
      m is not n
      every dead message in Q is m or n
then  Q.retry(m)
\end{lstlisting}

\paragraph{Incorrect scenario-inconsistency diagnosis.} In one of the runs of the third variant of \inlineconcept{NoDeadSensitiveMessages} at some point the model refused to produce a design, explaining that the scenarios were inconsistent and that no design could satisfy them. This occurred in the synthesis with 10 elicited scenarios that included the following ones.
\begin{lstlisting}[style=pacmplstyle, numbers=none, language=concepts]
ok:  Q.submit(m); Q.fail(m); L.affix(m, Sensitive); Q.purge();
nok: Q.submit(m); Q.fail(m); Q.submit(n); Q.fail(n); 
     L.affix(n, Sensitive); Q.purge();
\end{lstlisting}
The second scenario was the tenth to be elicited and the desired repair in that situation ended up not being elicited. At first glance, they indeed may seem inconsistent.

\paragraph{Redundant reactions and conditions.} A frequent quality failure in the synthesized designs was the inclusion of redundant reactions. This occurred in 15 out of the 82 runs that synthesized the intended design. For example, in one of the runs of the second variant of \inlineconcept{NoDeadSensitiveMessages} we got the following reaction.
\begin{lstlisting}[style=pacmplstyle, numbers=none, language=concepts]
reaction prevent_sensitive_dead_detach
when  L.detach(m, Sensitive)
where m is in dead in Q
      Sensitive is a label of m in L
then  error
\end{lstlisting}
In that variant, this is redundant because when a dead message is labelled as sensitive it is immediately purged by another reaction, so this reaction will actually never fire. Redundant reactions also occur frequently in cascading situations. For example, in one of the runs of the fourth variant of this example we got the following reactions.
\begin{lstlisting}[style=pacmplstyle, numbers=none, language=concepts]
reaction retry_other_dead_after_sensitive_affix
when  L.affix(s,Sensitive)
where s is in dead in Q
      m is in dead in Q
      m != s
      Sensitive is not a label of m in L
then  Q.retry(m)
reaction retry_next_other_dead
when  Q.retry(m)
where s is in dead in Q
      s != m
      Sensitive is a label of s in L
      n is in dead in Q
      n != m
      n != s
      Sensitive is not a label of n in L
then  Q.retry(n)
\end{lstlisting}
In this case, the second reaction is redundant because the first already triggers the retry of all non-sensitive dead messages at once.

Another very common situation (it occurs in almost every synthesized design) is the inclusion of redundant conditions, namely pre-conditions of the triggering action. One example of that occurs in the \inlineconcept{prevent_sensitive_dead_detach} reaction above, where the condition \inlineconcept{Sensitive is a label of m in L} is not needed, since that is already required by the precondition of \inlineconcept{L.detach(m, Sensitive)}.

Overall, the dominant correctness failures were not failures of bounded verification, but failures of specification coverage: missing scenarios and overfitting problems led the model to verified designs that satisfied the given invariant and scenarios but not the intended design.

\subsection{Threats to Validity}

\paragraph{Construct validity}

The plausibility classification used in RQ2 is necessarily partly subjective, because it concerns domain expectations that were not encoded as formal invariants. To mitigate this threat, we used plausibility only as a secondary measure: the main quantitative result of RQ2 is whether synthesis converges to verified designs and whether those designs are stable across runs. 

The assessment of whether a synthesized design was equivalent to the intended design was performed manually. To increase confidence, when applicable we checked the synthesized designs against the manually defined scenarios used in RQ3. However, this is not a complete equivalence check. A more robust evaluation would require automated refinement or equivalence checking between designs, which we leave for future work.

In RQ3, prompts and scenarios were revised after initial failures. This may introduce experimenter bias, since the revisions were informed by observed model behavior. We mitigate this by allowing at most one revision and by treating revised results separately from initial results. The revised runs should therefore be interpreted as modeling an interactive design process, not as one-shot synthesis performance.

\paragraph{Internal validity}

Our verification results also depend on the correctness of the translation from the concept and reaction descriptions to Alloy. Since invariants and reaction conditions are currently written in natural-language, \tool uses an LLM to translate them into Alloy. Translation errors could lead to false positives or false negatives: a correct design might be rejected, or an incorrect design might be accepted. 

We manually inspected most Alloy translations of the final verified synthesized designs and found no problems, but this inspection was not exhaustive. Translation errors may also have occurred in intermediate iterations. Such errors could mislead the counterexample-guided loop, although in our runs the synthesis usually still converged to a verified design; the only non-convergent case was one where the model judged the scenarios inconsistent. 

Moreover, ``verified'' means verified within the finite scope and step bound used by Alloy. The absence of a counterexample therefore does not imply unbounded correctness. This limitation is inherent to bounded model checking. In our evaluation, the chosen default scope and bound made verification fast enough to support interactive synthesis, but larger bounds may expose additional counterexamples.

\paragraph{External validity}

The main evaluation used GPT-5.5, so the results may not generalize to other LLMs. Different models may differ in their ability to synthesize reactions, generalize from scenarios, avoid overfitting, or translate conditions to Alloy. Since our technique treats the LLM largely as a replaceable component, we expect the verification loop to remain useful with other frontier models, although the number of iterations and failure modes may change. As a limited cross-model check, we repeated the scenario-guided synthesis experiment once for each benchmark variant using the initial scenarios and Claude Fable 5 with default parameters. Claude converged to the intended design for all twelve variants, doing so on the first attempt for six variants and requiring more than two attempts for only one. Although this additional run is insufficient for a systematic comparison between models, it provides initial evidence that the effectiveness of the synthesis procedure is not specific to GPT-5.5.


Our benchmark contains three applications and twelve design variants, which is not enough to represent the full diversity of concept-design problems. The variants were chosen to exercise different synchronization patterns, including error reactions, corrective reactions, cascading repairs, and alternative repair orders. Nevertheless, larger applications with more concepts, richer state, or many interacting invariants may expose scalability issues not observed here. In particular, verification cost grows quickly with scope, so applying the approach to larger designs may require modular verification or offline synthesis.

The benchmark variants were designed by the author, which may bias the evaluation toward designs that our technique handles well. We attempted to reduce this risk by including variants where the technique fails, such as cascading repairs and alternative repair orders, but independent benchmarks would provide stronger evidence.

\paragraph{Conclusion validity}

The conclusions are also limited by the number of runs. We ran synthesis three times per condition, which is enough to reveal several recurring failure modes but not enough for strong statistical claims. Scenario elicitation is also sensitive to non-determinism, and we generated only one batch of 10 scenarios and one batch of 20 scenarios per variant. As shown by the overlap results, different elicitation runs can explore different parts of the scenario space, so the reported success rates should be interpreted as indicative rather than definitive.

\section{Related Work}
\label{sec:relatedwork}

\paragraph{Concept Design} Concept Design was introduced as a methodology for structuring software around user-facing concepts~\cite{jackson2021essence}. In the original formulation, concepts were composed using simpler synchronization rules that did not include \emph{where} clauses. An early industry adoption report~\cite{wilczynski2023concept} focused mainly on the discovery and governance of individual concepts, with little detail on their composition. Other work has used Concept Design to analyze dark patterns~\cite{caragay2024beyond}, again focusing mainly on individual concepts and on mismatches between implemented concepts and users' expectations. Meng and Jackson introduced \emph{where} clauses and proposed an execution engine for concept synchronization rules~\cite{meng2025you}, also arguing that Concept Design makes software legible, including for LLMs generating implementations. Their synchronization rules allow sequences of actions in both \emph{when} and \emph{then} clauses; such rules can often be encoded in our simpler reaction form by introducing auxiliary reactions to chain actions and state predicates to record relevant past occurrences. Taking the legibility argument one step further, subsequent work proposed Concept Design as a mechanism for aligning meaning across all software artifacts and development activities~\cite{meng2026making}. That work coined the term ``reaction'' and used a high-level language for defining reactions similar to the one we use here. Although individual concepts have been modeled as state machines since the original presentation of Concept Design, prior work has not given a formal semantics for reactions or used such a semantics to verify and synthesize safe coordination logic. To our knowledge, ours is the first work to do so.

\paragraph{Scenario-based specification and synthesis} The use of scenarios as specification artifacts is far from new, being central in areas such as \emph{scenario-based design}~\cite{carroll2000making} or programming by demonstration~\cite{cypher1993watch}. Scenario-guided synthesis of behavioral models also has a long history. As early as 2000, Whittle and Schumann proposed a technique to generate UML statecharts from scenarios expressed as UML sequence diagrams~\cite{Whittle2000}. Uchitel et al. synthesized finite sequential processes from \emph{Message Sequence Charts} (MSCs)~\cite{1178048}. Later, Alur et al. synthesized finite-state distributed protocols modeled as finite-state I/O automata: their technique first generated incomplete automata from MSCs, and then completed those automata so that additional safety and liveness requirements were satisfied~\cite{DBLP:conf/hvc/AlurMRSTU14}. Although Concept Design reactions also have a semantics in terms of automata, we synthesize the high-level language that describes them directly. Another key difference is that we use an LLM, rather than a symbolic technique, to perform synthesis. The expressiveness of MSCs was extended in \emph{Live Sequence Charts} (LSCs)~\cite{DammHarel2001}, which can express mandatory and forbidden behavior and support execution through play-out, later becoming the basis of a classic scenario-based programming framework~\cite{HarelMarelly2003}. Note, however, that our scenarios do not define the full behavior of the application. They are positive and negative sequences of actions used to steer the synthesis of reactions. In this sense, LSCs are closer to Concept Design reactions themselves than to the scenarios used in our synthesis loop.

\paragraph{LLMs for High-Level Software Design} The use of LLMs in software engineering is now widespread, being applied in most phases of software development~\cite{hou2024large}. Their use in architecture and design tasks is also increasing, although recent surveys report that conformance checking and rigorous validation of generated artifacts remain underexplored~\cite{schmid2025software,esposito2025generative}. Automated generation of design models from natural-language requirements predates LLMs, with earlier NLP-based approaches surveyed by Ahmed et al.~\cite{ahmed2022automatic}. Recent LLM-based work revisits this problem by generating UML class diagrams from user stories or requirements~\cite{10633327,giannouris2025nomad}, and UML sequence diagrams from requirements~\cite{ferrari2024model}. Sequence-diagram generation is related to our scenario elicitation problem, because sequence diagrams are scenario-like behavioral artifacts. However, our goal is not to generate scenarios or diagrams as final design artifacts, but to use classified scenarios to synthesize general coordination rules.

\paragraph{LLM-Driven Synthesis and Autoformalization} Program synthesis has been one of the most successful applications of LLMs in software engineering~\cite{austin2021program, chen2021evaluating}. Our synthesis loop follows the classical CEGIS pattern~\cite{solar2008program}, in which a candidate generator is paired with a verifier that returns counterexamples for refinement. Recent work has also used counterexample or property feedback to improve LLM-based synthesis~\cite{pereira2026property,orvalho2025counterexample}. Our work differs in applying this pattern to high-level coordination-design synthesis, and in using scenarios as checkable steering artifacts for intended behavior rather than only verifying candidates against a formal correctness property. Although not central to our technique, \tool also uses an LLM to translate concept designs expressed in a language with natural-language elements into Alloy for formal verification. This aligns with autoformalization, the translation of informal statements into verifiable formal representations~\cite{weng2025autoformalization}. The generally successful Alloy translations in our benchmark are consistent with prior work on LLM-assisted formalization and repair of declarative specifications, including work targeting Alloy~\cite{alhanahnah2025empirical, hong2025effectiveness}.

\section{Conclusion}
\label{sec:conclusion}

This work provides a formal semantics for designs developed with the Concept Design methodology, enabling bounded checking of both invariants and scenarios. Building on this semantics, \tool implements a CEGIS-style LLM-driven synthesis loop for reaction designs. Our evaluation confirms that verification is essential: LLM-generated designs can often satisfy a simple invariant, but still fail to do so occasionally. More importantly, invariants alone often leave the intended coordination behavior underconstrained. Scenarios provide a checkable steering artifact: they guide synthesis more reliably than natural-language prompts in our benchmark, are easier to revise mechanically, and can be elicited using an LLM. At the same time, the results show that scenario coverage matters; missing or overly minimal scenarios can still lead to verified but unintended designs.

Our long-term goal is to make verified synthesis practical for realistic Concept Design applications. Several limitations of the current prototype point to the next steps. First, the translation to Alloy should be made trustworthy by replacing LLM-based translation with a formal source language for concepts, reactions, invariants, and scenarios, together with a symbolic compiler to the verification backend. The challenge is to make that language precise enough for verification while preserving the legible, natural-language style that makes Concept Design usable and LLM-friendly. Second, verification must scale beyond the small bounded instances used in this evaluation. This will likely require more efficient encodings, modular verification techniques, or offline synthesis workflows for larger applications. Third, users need better support for comparing competing designs, so we plan to add automated refinement or equivalence checking between reaction designs to \tool. Scenario elicitation also remains an open design problem: future protocols should combine LLM exploration with symbolic enumeration of invariant-violating prefixes, so that elicitation covers both diverse violations and alternative repairs for the same violation. Finally, a more extensive evaluation should address the external and conclusion validity threats discussed above, using more third-party applications and design variants, more LLMs, and more runs per experiment to account for nondeterminism.

\section{Acknowledgements}

The author would like to thank the members of the Software Design Group at MIT, namely Abutalib (Barish) Namazov, Carmel Schare, Daniel Jackson, and Eagon Meng for all the fruitful discussions that influenced the development of this work.
This work was financed by National Funds through the FCT - Fundação para a Ciência e a Tecnologia, I.P. under the PRR - Recovery and Resilience Plan, within the scope of the Science + Training measure, within grant FCT/Mobility/1355087736/2024-25. GPT-5.5 was used to review the text, draw the figures, and improve some of the tables. The threats to validity section was directly written by GPT-5.5 following an itemized summary provided by the author.  

\bibliography{references}

@book{jackson2021essence,
  title={The Essence of Software: Why Concepts Matter for Great Design},
  author={Jackson, Daniel},
  publisher={Princeton University Press},
  year={2021},
  address={Princeton, NJ},
  isbn={9780691225388}
}

@inproceedings{meng2025you,
  title={What You See Is What It Does: A Structural Pattern for Legible Software},
  author={Meng, Eagon and Jackson, Daniel},
  booktitle={Proceedings of the 2025 ACM SIGPLAN International Symposium on New Ideas, New Paradigms, and Reflections on Programming and Software},
  pages={178--193},
  year={2025},
  publisher={Association for Computing Machinery},
  address={New York, NY, USA}
}

@article{meng2026making,
  title={Making Software Meaningful},
  author={Meng, Eagon and Namazov, Abutalib and Schare, Carmel and Cunha, Alcino and Jackson, Daniel},
  journal={arXiv preprint arXiv:2606.11051},
  year={2026}
}

@book{jackson2012software,
  title={Software Abstractions: logic, language, and analysis},
  author={Jackson, Daniel},
  year={2012},
  publisher={The MIT Press},
  edition={2nd}
}

@inproceedings{macedo2016lightweight,
  title={Lightweight specification and analysis of dynamic systems with rich configurations},
  author={Macedo, Nuno and Brunel, Julien and Chemouil, David and Cunha, Alcino and Kuperberg, Denis},
  booktitle={Proceedings of the 2016 24th ACM SIGSOFT International Symposium on Foundations of Software Engineering},
  pages={373--383},
  year={2016}
}

@inproceedings{brunel2018electrum,
  title={The electrum analyzer: model checking relational first-order temporal specifications},
  author={Brunel, Julien and Chemouil, David and Cunha, Alcino and Macedo, Nuno},
  booktitle={Proceedings of the 33rd ACM/IEEE International Conference on Automated Software Engineering},
  pages={884--887},
  year={2018}
}

@article{hong2025effectiveness,
  title={On the effectiveness of large language models in writing {Alloy} formulas},
  author={Hong, Yang and Jiang, Shan and Fu, Yulei and Khurshid, Sarfraz},
  journal={arXiv preprint arXiv:2502.15441},
  year={2025}
}

@article{alhanahnah2025empirical,
  title={An empirical evaluation of pre-trained large language models for repairing declarative formal specifications},
  author={Alhanahnah, Mohannad and Rashedul Hasan, Md and Xu, Lisong and Bagheri, Hamid},
  journal={Empirical Software Engineering},
  volume={30},
  number={5},
  pages={149},
  year={2025},
  publisher={Springer}
}

@article{lamport1994temporal,
  title={The temporal logic of actions},
  author={Lamport, Leslie},
  journal={ACM Transactions on Programming Languages and Systems (TOPLAS)},
  volume={16},
  number={3},
  pages={872--923},
  year={1994},
  publisher={ACM New York, NY, USA}
}

@inproceedings{caragay2024beyond,
  title={Beyond dark patterns: A concept-based framework for ethical software design},
  author={Caragay, Evan and Xiong, Katherine and Zong, Jonathan and Jackson, Daniel},
  booktitle={Proceedings of the 2024 CHI Conference on Human Factors in Computing Systems},
  pages={1--16},
  year={2024}
}

@inproceedings{wilczynski2023concept,
  title={Concept-centric software development: An experience report},
  author={Wilczynski, Peter and Gregoire-Wright, Taylor and Jackson, Daniel},
  booktitle={Proceedings of the 2023 ACM SIGPLAN International Symposium on New Ideas, New Paradigms, and Reflections on Programming and Software},
  pages={120--135},
  year={2023}
}

@inproceedings{DBLP:conf/hvc/AlurMRSTU14,
  author       = {Rajeev Alur and
                  Milo M. K. Martin and
                  Mukund Raghothaman and
                  Christos Stergiou and
                  Stavros Tripakis and
                  Abhishek Udupa},
  title        = {Synthesizing Finite-State Protocols from Scenarios and Requirements},
  booktitle    = {Haifa Verification Conference},
  series       = {Lecture Notes in Computer Science},
  volume       = {8855},
  pages        = {75--91},
  publisher    = {Springer},
  year         = {2014}
}

@ARTICLE{1178048,
  author={Uchitel, S. and Kramer, J. and Magee, J.},
  journal={IEEE Transactions on Software Engineering}, 
  title={Synthesis of behavioral models from scenarios}, 
  year={2003},
  volume={29},
  number={2},
  pages={99-115},
  doi={10.1109/TSE.2003.1178048}}

@book{carroll2000making,
  title     = {Making Use: Scenario-Based Design of Human-Computer Interactions},
  author    = {Carroll, John M.},
  year      = {2000},
  publisher = {The MIT Press},
  address   = {Cambridge, MA, USA},
  isbn      = {9780262032797}
}

@book{cypher1993watch,
  title={Watch What I Do: Programming by Demonstration},
  author={Cypher, Allen},
  year={1993},
  publisher={MIT Press},
  address={Cambridge, MA, USA}
}

@inproceedings{Whittle2000,
  author    = {Jon Whittle and Johann Schumann},
  title     = {Generating statechart designs from scenarios},
  booktitle = {Proceedings of the 22nd International Conference on Software Engineering (ICSE)},
  pages     = {314--323},
  year      = {2000},
  doi       = {10.1145/337180.337217},
  url       = {https://dl.acm.org/doi/10.1145/337180.337217}
}

@article{DammHarel2001,
  author    = {Werner Damm and David Harel},
  title     = {LSCs: Breathing life into message sequence charts},
  journal   = {Formal Methods in System Design},
  year      = {2001},
  volume    = {19},
  number    = {1},
  pages     = {45--80},
  doi       = {10.1023/A:1011227529550},
  url       = {https://doi.org}
}

@book{HarelMarelly2003,
author = {Harel, David and Marelly, Rami},
title = {Come, Let's Play: Scenario-Based Programming Using LSC's and the Play-Engine},
year = {2003},
isbn = {3540007873},
publisher = {Springer-Verlag},
address = {Berlin, Heidelberg}
}

@article{hou2024large,
  title={Large language models for software engineering: A systematic literature review},
  author={Hou, Xinyi and Zhao, Yanjie and Liu, Yue and Yang, Zhou and Wang, Kailong and Li, Li and Luo, Xiapu and Lo, David and Grundy, John and Wang, Haoyu},
  journal={ACM Transactions on Software Engineering and Methodology},
  volume={33},
  number={8},
  pages={1--79},
  year={2024},
  publisher={ACM New York, NY}
}

@article{schmid2025software,
  title={Software architecture meets {LLMs}: A systematic literature review},
  author={Schmid, Larissa and Hey, Tobias and Armbruster, Martin and Corallo, Sophie and Fuch{\ss}, Dominik and Keim, Jan and Liu, Haoyu and Koziolek, Anne},
  journal={arXiv preprint arXiv:2505.16697},
  year={2025}
}

@INPROCEEDINGS{10633327,
  author={Li, Yishu and Keung, Jacky and Ma, Xiaoxue and Chong, Chun Yong and Zhang, Jingyu and Liao, Yihan},
  booktitle={2024 IEEE 48th Annual Computers, Software, and Applications Conference (COMPSAC)}, 
  title={LLM-Based Class Diagram Derivation from User Stories with Chain-of-Thought Promptings}, 
  year={2024},
  volume={},
  number={},
  pages={45-50},
  doi={10.1109/COMPSAC61105.2024.00017}}

@inproceedings{ferrari2024model,
  title={Model generation with {LLMs}: From requirements to UML sequence diagrams},
  author={Ferrari, Alessio and Abualhaija, Sallam and Arora, Chetan},
  booktitle={2024 IEEE 32nd International Requirements Engineering Conference Workshops (REW)},
  pages={291--300},
  year={2024},
  organization={IEEE}
}

@article{esposito2025generative,
  title={Generative {AI} for software architecture. Applications, challenges, and future directions},
  author={Esposito, Matteo and Li, Xiaozhou and Moreschini, Sergio and Ahmad, Noman and Cerny, Tomas and Vaidhyanathan, Karthik and Lenarduzzi, Valentina and Taibi, Davide},
  journal={Journal of Systems and Software},
  pages={112607},
  year={2025},
  publisher={Elsevier}
}

@inproceedings{ahmed2022automatic,
  title={Automatic transformation of natural to unified modeling language: A systematic review},
  author={Ahmed, Sharif and Ahmed, Arif and Eisty, Nasir U},
  booktitle={2022 IEEE/ACIS 20th International Conference on Software Engineering Research, Management and Applications (SERA)},
  pages={112--119},
  year={2022},
  organization={IEEE}
}

@article{giannouris2025nomad,
  title={{NOMAD}: A Multi-Agent {LLM} System for {UML} Class Diagram Generation from Natural Language Requirements},
  author={Giannouris, Polydoros and Ananiadou, Sophia},
  journal={arXiv preprint arXiv:2511.22409},
  year={2025}
}

@book{solar2008program,
  title={Program synthesis by sketching},
  author={Solar-Lezama, Armando},
  year={2008},
  publisher={University of California, Berkeley}
}

@article{chen2021evaluating,
  title={Evaluating large language models trained on code},
  author={Chen, Mark and Tworek, Jerry and Jun, Heewoo and Yuan, Qiming and Pinto, Henrique Ponde De Oliveira and Kaplan, Jared and Edwards, Harri and Burda, Yuri and Joseph, Nicholas and Brockman, Greg and others},
  journal={arXiv preprint arXiv:2107.03374},
  year={2021}
}

@article{austin2021program,
  title={Program synthesis with large language models},
  author={Austin, Jacob and Odena, Augustus and Nye, Maxwell and Bosma, Maarten and Michalewski, Henryk and Dohan, David and Jiang, Ellen and Cai, Carrie and Terry, Michael and Le, Quoc and others},
  journal={arXiv preprint arXiv:2108.07732},
  year={2021}
}

@article{pereira2026property,
  title={Property-Guided LLM Program Synthesis for Planning},
  author={Pereira, Andr{\'e} G and Corr{\^e}a, Augusto B and Seipp, Jendrik},
  journal={arXiv preprint arXiv:2605.16142},
  year={2026}
}

@inproceedings{orvalho2025counterexample,
  title={Counterexample guided program repair using zero-shot learning and maxsat-based fault localization},
  author={Orvalho, Pedro and Janota, Mikol{\'a}{\v{s}} and Manquinho, Vasco M},
  booktitle={Proceedings of the AAAI Conference on Artificial Intelligence},
  volume={39},
  number={1},
  pages={649--657},
  year={2025}
}

@article{weng2025autoformalization,
  title={Autoformalization in the era of large language models: A survey},
  author={Weng, Ke and Du, Lun and Li, Sirui and Lu, Wangyue and Sun, Haozhe and Liu, Hengyu and Zhang, Tiancheng},
  journal={arXiv preprint arXiv:2505.23486},
  year={2025}
}

\end{document}